\documentclass[11.5pt]{article}
\usepackage{}
\usepackage{wrapfig}
\usepackage{cite,amssymb,amsfonts,hangcaption,amsmath,here,color}
\usepackage{graphicx,here}

\renewcommand{\theequation}{\arabic{section}.\arabic{equation}}

\makeatletter
\@addtoreset{equation}{section}
\@addtoreset{footnote}{section}
\makeatother

\makeatletter
\renewcommand\appendix{
\par
\setcounter{section}{0}%
\setcounter{subsection}{0}%
\gdef\thesection{Appendix \@Alph\c@section }
\renewcommand{\theequation}
{\Alph{section}.\arabic{equation}}
}
\makeatother

\makeatletter
\def\eqnarray{ \stepcounter{equation} \let\@currentlabel=\theequation
\global\@eqnswtrue
\global\@eqcnt\z@
\tabskip\@centering
\let\\=\@eqncr
$$\halign to \displaywidth\bgroup\@eqnsel\hskip\@centering
$\displaystyle\tabskip\z@{##}$&\global\@eqcnt\@ne
\hfil$\displaystyle{{}##{}}$\hfil
&\global\@eqcnt\tw@$\displaystyle\tabskip\z@{##}$\hfil
\tabskip\@centering&\llap{##}\tabskip\z@\cr}
\makeatother

\makeatletter
\def\@arrayacol{\edef\@preamble{\@preamble \hskip .5\arraycolsep}}
\def\array{\let\@acol\@arrayacol \let\@classz\@arrayclassz
\let\@classiv\@arrayclassiv \let\\\@arraycr\def\@halignto{}\@tabarray}
\makeatother

\makeatletter
\newcounter{subeqncnt}
\def\thesubeqncnt{\alph{subeqncnt}}
\def\subequations{\begingroup%
\stepcounter{equation}\edef\@tempa{\theequation}%
\let\c@equation\c@subeqncnt\c@subeqncnt\z@
\edef\theequation{\@tempa\noexpand\thesubeqncnt}}

\makeatother

\newcommand{\captionfonts}{\small}
\makeatletter 
\long\def\@makecaption#1#2{%
\vskip\abovecaptionskip
\sbox\@tempboxa{{\captionfonts #1: #2}}%
\ifdim \wd\@tempboxa >\hsize
{\captionfonts #1: #2\par}
\else
\hbox to\hsize{\hfil\box\@tempboxa\hfil}%
\fi
\vskip\belowcaptionskip}
\makeatother 

\def\ba{\begin{eqnarray}}
\def\ea{\end{eqnarray}}

\begin{document}

\setlength{\baselineskip}{7mm}
\begin{titlepage}
\begin{flushright}
{\tt CAS-KITPC/ITP-269} \\
{\tt July 2011} \\
\end{flushright}



\setlength{\baselineskip}{9mm}

\begin{center}

\vspace*{15mm}
{\LARGE
Quark Number Susceptibility and QCD Phase Transition in the Predictive Soft-wall AdS/QCD Model with Finite Temperature}

\vspace{8mm}
{\large
{\sc{Ling-Xiao Cui}}
,
{\sc{Shingo Takeuchi}}
 and
{\sc{Yue-Liang Wu}}
}

\setlength{\baselineskip}{0mm}

\vspace*{12mm}
{\large
{\it{Kavli Institute for Theoretical physics China (KITPC)}}\\
{\it{Key Laboratory of Frontiers in Theoretical Physics}}\\
{\it{Institute of Theoretical Physics, Chinese Academy of Sciences, Beijing 100190, China}} \\
}

\end{center}

\vspace{1cm}

{\large
\begin{abstract}
Based on the infrared improved soft-wall AdS/QCD model which can lead to a consistent prediction for the mass spectra of light resonance mesons, we extend it to the finite temperature system and carry out the calculation for the quark number susceptibility with finite quark mass.  As a consequence, we show that the quark number susceptibility grows rapidly with a continuous blow-up in a narrow
temperature interval as the temperature increases, and there is a peak when the temperature is around $T \simeq 160 \sim 190$ MeV depending slightly on the models, the resulting critical temperature is about 170MeV, which agrees remarkably with the lattice QCD calculation.

\end{abstract}
}

\end{titlepage}
\section{Introduction}

Quantum chromodynamics describes the dynamics of quarks.
Its gauge group is non-Abelian and the gauged gluon fields have self-interactions.
As a result, the beta-function at ultraviolet (UV) region becomes negative that leads to the asymptotic freedom \cite{Gross:1973id,Politzer:1973fx}.
While at low energies, the non-perturbative dynamics such as chiral symmetry breaking and linear confinement appear as the gauge coupling constant grows.

The duality between gravities and gauge theories \cite{Maldacena:1997re,Gubser:1998bc,Witten:1998qj} gives a new approach to the study of
the low energy QCD. Among the studies of the duality, the ones developed phenomenologically are referred as bottom-up approach
(AdS/QCD model) \cite{Erlich:2005qh,DP} (There are also good reviews, e.g.\cite{reviewAdSQCD}). One of the important achievements in the AdS/QCD model are the realization of
the chiral symmetry breaking in the hard-wall AdS/QCD model \cite{Erlich:2005qh} and the linear confinement in the soft-wall
AdS/QCD model\cite{Karch:2006pv}. But in the hard-wall model, the resulting excited meson mass spectra are found to be contrary to the experiments.
In the soft-wall model, one can obtain desired mass spectra for the excited vector mesons, but the chiral symmetry breaking occurs inconsistently.
Several groups studied on these problems\cite{Colangelo:2008us,Gherghetta:2009ac}. A predictive soft-wall AdS/QCD model was constructed by simply modifying the 5D AdS metric in the infrared (IR) region\cite{Sui:2009xe} and the resulting mass spectra for the light resonance states of pseudoscalar, scalar, vector and axial-vector mesons remarkably agree with the experimental data. It is then interesting to study the phase transition for such a holographic AdS/QCD model with finite temperature, which comes to the purpose of our present paper.  Thus we will extend such an infrared improved soft-wall AdS/QCD model\cite{Sui:2009xe} to a finite temperature system by considering a black hole metric and investigate the phase transition by computing the quark number susceptibility (QNS).
%
The QNS is known to be the response of quark number density to infinitesimal change of the quark chemical potential as $\chi_q \equiv {d n_q}/{d \mu_q}$. Using the fluctuation-dissipation theorem, it can be expressed as \cite{Kunihiro:91}
\begin{equation}\label{suss}
\chi_q = -\lim_{k\to 0} {\mbox{Re}}~G^{\rm R}_{t\ t}(\omega=0, k),
\end{equation}
where $G^{\rm R}_{t \ t}(\omega, k)$ is the two-point retarded Green function for time-component of quark number current.
%
From the lattice QCD studies\cite{Mc87, AlltonEjiri}, it strongly suggests that the behavior of $\chi_q/T^2$ on the transition between the hadronic phase and the Quark-Gluon-Plasma phase at zero quark chemical potential is a continuous but rapid in a narrow
temperature interval around $T \simeq 170$ MeV \cite{review}.
%
In this paper, we are going to compute the QNS with the constituent quark mass $\bar{m}_q$ due to the dynamically generated chiral condensation of quark $\sigma=<\bar{q}q>$, which is defined as $\bar{m}_q = m_q + \sigma/\mu^2$ with $m_q$ the current quark mass and $\sigma/\mu^2$ regarded as the dynamically generated quark mass for the typical QCD scale $\mu\sim \Lambda_{QCD}$ due to QCD confinement\cite{DW}, and to show how the behavior of $\chi_q/T^2$ changes as the temperature and dynamical quark mass and what is the critical temperature
at zero chemical potential.
%

This paper is organized as follows: in Sect.\ref{Chap:The model}, we present an introduction to our working model and the $U(1)$ gauge field dual to the quark number current. In Sect.\ref{Chap:The result}, we show how to get the quark number susceptibility from the solution of the $U(1)$ gauge field, and provide the numerical results with a detailed discussion. Our summary and remarks are presented in Sect.\ref{chap:Summary}.

\section{The models}\label{Chap:The model}

As the predictive soft-wall AdS/QCD model constructed by simply modifying the 5D AdS metric in the infrared (IR) region\cite{Sui:2009xe} can result the consistent mass spectra for the light resonance states of pseudoscalar, scalar, vector and axial-vector mesons, we shall extent such a model
to a finite temperature system for studying the phase transition. This may simply be realized by considering a 5D black hole metric as follows
\begin{equation}\label{metric_dads_01}
ds^2 = a^2(z) \left( f(z) dt^2 + \sum_{i=1}^3 dx^2_i - \frac{dz^2}{f(z)} \right),
\end{equation}
with
\begin{eqnarray}\label{f(z)}
a^2(z) = 1/z^2 + \mu_g^2  \quad {\rm and} \quad f(z) &=& 1-(z/z_0)^4,
\end{eqnarray}
where $z$ has the relation with the normal radial coordinate $r$ as $z \equiv 1/r$ and $z_0$ is the horizon.
The scale factor $\mu_g$ characterizes the confinement of QCD in the infrared region by breaking the conformal symmetry, it is fixed by fitting with a linear function for the excited states and has been given in \cite{Sui:2009xe} from a consistent prediction for the resonance messon mass spectra. Hawking temperature is given as $T = 1/(z_0\pi)$.
%

The action with the background dilaton $\Phi(z)$ which has a nontrivial solution for forming the soft-wall is given as \cite{Sui:2009xe}
\begin{eqnarray}\label{lagran}
S_\Phi=\int d^5x\sqrt{g}e^{-\Phi(z)}\,{\rm{Tr}} \Big( |D_M X(x,z)|^2-m_X^2 |X(x,z)|^2 \Big),
\end{eqnarray}
where $m_X^2 = -3$ so that the solution of dilaton $\Phi(z)$ obtained below does not have the singular at UV-limit ($z \to 0$).
%
%
$X(x,z)$ is the bulk scalar field, and we denote its vacuum expectation value (VEV) as $X(z)$ for simplicity
\begin{eqnarray} \label{xexpect0}
X(z) = \frac{1}{2}\,v(z)~\mathbf{1}_2,
\end{eqnarray}
where $\mathbf{1}_2$ denotes $2 \times 2$ unit matrix.
For our purpose in the present note, we consider that the VEV $X(z)$ is taken as the same form given in \cite{Sui:2009xe}, which is listed in Table.\ref{formsofv}.
While, the dilaton $\Phi(z)$ is fixed from the equation of motion for $X(z)$ which is given as
\begin{eqnarray}\label{xexpect}
0 = \partial_z \Big( a^3(z) f(z) e^{-\Phi(z)} \partial_z v(z) \Big) - a^5(z) e^{-\Phi(z)} m_X^2 v(z).
\end{eqnarray}
where the resulting $\Phi(z)$ from eq.(\ref{xexpect}) will be different from the one yielded in \cite{Sui:2009xe} due to the inclusion of 
the black hole metric $f(z)$.
%
%
%
%
%
\begin{table}[H]
\begin{center} \begin{tabular}{llll}
\hline\hline
Model & $\qquad\qquad$ $v(z)$ & \qquad Parameters & \\
\hline
\quad Ia  & $ z(A+Bz^2)(1+Cz^2)^{-1} $  & \quad $B=\sigma/{\zeta}+m_q\zeta C$, & \quad  $C=B / (\mu_d \gamma) $\\
\quad Ib  & $ z(A+Bz^2)(1+Cz^2)^{-5/4}$ & \quad $B=\sigma/{\zeta}+\frac{5}{4}m_q\zeta C$, & \quad  $C=(B^2/ (\mu_d \gamma^2))^{2/5}$ \\
\quad IIa & $ z(A+Bz^2)(1+Cz^4)^{-1/2}$ & \quad $B=\sigma/{\zeta}$, & \quad  $C=(B/( \mu_d\gamma))^2$ \\
\quad IIb & $ z(A+Bz^2)(1+Cz^4)^{-5/8}$ & \quad $B=\sigma/{\zeta}$, & \quad  $C=(B^2/(\mu_d\gamma^2))^{4/5}$ \\
\hline\hline
\end{tabular}
\caption{ $A=m_q\zeta$ and $\zeta=\sqrt{3}/(2\pi)$ for all the models, while $\mu_d^2 \equiv 3/(2\alpha)\mu_g^2$ with $\alpha = 1$
for the Ia, IIa and $\alpha = 1/2$ for the Ib, IIb. $A,B$ and $C$ satisfy eq.(\ref{vzero}).
Numerical values of $m_q$,$\gamma$ and $\sigma$ are given in Table.\ref{parameter}.}\label{formsofv}
\end{center}
\end{table} %
%
 Such an VEV may be regarded as the one at the low temperature phase. It was shown that the boundary behaviors of the VEV play an essential role\cite{Sui:2009xe} for mass spectra of resonances mesons. In general, its UV boundary behavior is fixed by the AdS/CFT correspondence, while its IR boundary behavior is imposed so as to obtain the required boundary behavior of dilaton at the IR-region. From Table.1, we can see that $v(z)$ behaves in the IR-limit $z\to \infty$ and the UV-limit $z\to 0$ as
\begin{eqnarray}
v(z)\big|_{z \rightarrow \infty} &=& \gamma \left( \sqrt{\frac{3}{2\alpha}} \mu_g z \right)^\alpha \equiv \gamma \left( \mu_d z\right)^\alpha, \label{vinft} \\
v(z)\big|_{\rm UV-limit} &=& m_q \,\zeta\, z+\frac{\sigma}{\zeta}z^3. \label{vzero}
\end{eqnarray}
where $\mu_d^2$ is defined with $\alpha = 1$ for the Model Ia and IIa and $\alpha = 1/2$ for the Model Ib and IIb, and $\zeta$ is fixed as $\zeta=\sqrt{3}/(2\pi)$ \cite{Sui:2009xe}.
$m_q$ and $\sigma$ denote the current quark mass and the chiral condensation respectively, and the constituent quark mass in this paper is defined as  $\bar{m}_q = m_q + \sigma/\mu^2$\cite{DW}. Their numerical values are given in Table.\ref{parameter}, which have been fixed by minimizing the breaking of the Gell-Mann-Oakes-Renner relation as
\begin{eqnarray}\label{GMOR}
f_{\pi}^2~m_{\pi}^2 = 2m_q~\sigma
\end{eqnarray}
at the $1\%$ level and the experimental values: $m_\pi=139.6$ MeV and $f_\pi=92.4$ MeV.

\begin{table}[ht!]
\begin{center}
\begin{tabular}{ccccc}
\hline\hline
Model & Ia & Ib & IIa & IIb \\
\hline
$m_q$ (MeV)                  & 4.16  & 4.64  & 4.44  &  4.07\\
$\sigma^{\frac{1}{3}}$ (MeV) & 275   & 265   & 265   &  272\\
$\gamma$                     & 0.178 & 0.136 & 0.153 &  0.112\\
\hline\hline
\end{tabular}
\caption{$\mu_d^2 \equiv 3/(2\alpha)\mu_g^2=445^2$ with $\alpha = 1$ for the Ia, IIa and $\alpha = 1/2$ for the Ib, IIb.}
\label{parameter}
\end{center}
\end{table}

To compute QNS by using eq.(\ref{suss}), let us now introduce an $U(1)$ gauge field which is dual to the $4$D quark number current. 
Changing the coordinate to $u\equiv (z/z_H)^2$, and considering the following action
\begin{equation}\label{actionF2}
S = -\frac{1}{4g_5^2} \int du d^4x \sqrt{g} e^{-\Phi(u)} F_{mn} F^{mn}.
\end{equation}
we can simply get the equation of motion $ \displaystyle   0 = \partial_\mu (\sqrt{g} e^{-\Phi} F^{\mu\nu})$ for  $A_\nu$.
Performing the Fourier transformation as $A_\mu(t, x, u)= \int\!\frac{d^4 k}{(2\pi)^4} \ e^{-iw t+i x k}\tilde{A}_\mu(\omega, k, u)$, choosing the momenta $k$ along the $x$-direction and taking the axial gauge $A_u(u)=0$, we obtain the
following equations of motion for the needed components
\begin{eqnarray}\label{eqm3}
A_t ~:~ 0 &=& \left(e^{-\Phi} \sqrt{g} g^{uu} g^{tt} A_t' \right)' - q^2 e^{-\Phi}\sqrt{g} g^{xx} g^{tt} A_t - \omega q e^{-\Phi} \sqrt{g}  g^{xx} g^{tt} A_z, \\
A_u ~:~ 0 &=& \omega g^{tt} A_t' - q g^{xx} A_z',
\end{eqnarray}
where $q$ and $\omega$ are defined as $q\equiv k/(2\pi T)$ and $\omega \equiv w /(2\pi T)$, the prime denotes derivative with regard to $u$,
and in this study we use  $x$-direction as a usual $z$-direction in four dimensional space-time due to what we have already used the symbol $z$ as the fifth dimension here. In the above equations, we represent $\tilde{A}_\mu$ as $A_\mu$, and will apply this manner in this paper for simplicity.

\section{The results}\label{Chap:The result}

It turns out that the solution $\Phi(u)$ can be obtained analytically from the equation of motion eq.(\ref{xexpect}) for all the
models\footnote{We use mathematica 8.}.
In obtaining the solution for the dilaton $\Phi(u)$, we take the integral constant so that the solution at the boundary behaves as
\begin{eqnarray}\label{dilaton}
\Phi(z) \big|_{\rm UV-limit} =3 \mu_g^2 z^2,
\end{eqnarray}
Here we abbreviate to
represent its explicit results due to a long expression. By using
the solutions $\Phi(u)$ in all the models and the equation of motion
eq.(\ref{eqm3}), we are able to get the retarded green function.

From eq.(\ref{eqm3}), we can obtain the following equation
\begin{eqnarray}
0 =
\left( \frac{\psi'}{e^{-\Phi} \sqrt{g}  g^{xx} g^{tt}} \right)'
- \left( \frac{\omega^2 g^{tt}/g^{xx} + q^2}{e^{-\Phi} \sqrt{g}  g^{uu} g^{tt}} \right)\psi
\end{eqnarray}
with $\psi \equiv e^{-\Phi} \sqrt{g} g^{uu} g^{tt} A_t'$. Note that
there is a singular at $u=1$ due to the horizon, and the deepest one is with the power $-2$. To get rid of such a singular,
we assume that the solution of $\psi$ has the form $\psi = (1-u)^{i\nu\omega} X(u,\omega,q) $ with the regular function $X(u,\omega,q)$.
Then by appropriately determining the index $\nu$, we can regularize the singular behavior.
In general, there are two options for the values of $\nu$, which corresponds to the in-going and out-going solutions respectively.
In our present consideration, as the value of $\omega$ is taken to be zero eventually following eq.(\ref{suss}), it is not important to specify which one we are actually taking.
In the following treatments, we consider the case that $\omega, q \ll 1$ in the hydrodynamic expansion.
With such an assumption, the form of the function $X(u,\omega,q)$ can be expressed in the general form: 
$X(u,\omega,q)=\psi_0 + q^2 \psi_1(u) + \omega^2 \psi_2(u)+{\cal O}(\omega^4,q^4)$.
As a result, provided that $\nu \to 0$ for $\omega \to 0$, we can show that $\psi_0$ and $\psi_1$ satisfy the following equations
%
%
\begin{eqnarray}
0 &=& \left(\frac{\psi_0'}{e^{-\Phi}\sqrt{g} g^{xx} g^{tt}}\right)',  \\
0 &=& \left(\frac{\psi_1'}{e^{-\Phi}\sqrt{g} g^{xx} g^{tt}}\right)'  -  \frac{\psi_0}{e^{-\Phi}\sqrt{g} g^{uu} g^{tt}}.
\end{eqnarray}
It is then not difficult to find out the solution for $\psi_0$
\begin{eqnarray}
\psi_0 = C_1 + C_2 \int_1^u du e^{-\Phi}\sqrt{g}g^{xx} g^{tt}.
\end{eqnarray}
as the integrand $\sqrt{g}g^{xx}g^{tt} $ has the singularity arising from the horizon, we then take $C_2=0$. And the solution for
$\psi_1$ is given
\begin{eqnarray}
\psi'_1 &=& C_1  e^{-\Phi}\sqrt{g}g^{xx}g^{tt}
\left(
\int_1^u du' \frac{1}{e^{-\Phi}\sqrt{g}g^{uu}g^{tt}} + D_1
\right),
\end{eqnarray}
From the condition $\psi'(u=1)=0$, the integral constant $D_1$ can
be fixed. From the $A_t-$component of eq.(\ref{eqm3}) and the
definition of $\psi$ with $\omega \to 0$, we have
\begin{eqnarray}
A_t = C_1
\left(
\int_1^u du' \frac{1}{e^{-\Phi}\sqrt{g}g^{uu}g^{tt}} + D_1
\right)
+ {\cal O}(\omega^2,q^2).
\end{eqnarray}
where $C_1$ can be fixed by using $\displaystyle \lim_{u \to 0} A_t
\equiv A^{(0)}_t$ to be
\begin{eqnarray}
C_1 =
A^{(0)}_t \left( \int^0_1 du' \frac{1}{e^{-\Phi} \sqrt{g}g^{uu} g^{tt}} + D_1 \right)^{-1} + {\cal O}(\omega^2,q^2).
\end{eqnarray}
Here, $A^{(0)}_t$ plays the role as the source for the 4D quark number current mentioned in the above eq.(\ref{actionF2}).
From the definition of $\psi$, we can obtain the following expression
\begin{eqnarray}
A_t' = \frac{C_1}{e^{-\Phi}\sqrt{g}g^{uu}g^{tt}} + {\cal O}(\omega^2,q^2) + {\cal O}(\omega^2,q^2).
\end{eqnarray}

We now come to discuss the retarded green function through the GKP-W
relation \cite{Gubser:1998bc,Witten:1998qj} and the prescription
\cite{Policastro:2002se}. It turns out that the on-shell action
takes
\begin{equation}
S_0 =
\frac{e^{-\Phi} \sqrt{g}g^{uu}g^{tt}}{2g_5^2}
\int \frac{d^4k}{(2\pi)^4}
\left(  A_t A_t'- \frac{1}{f} A_x A'_x \right) \bigg|^0_{u=1}.
\end{equation}
Thus the retarded green function $G^{\rm R}_{t~t}$ is found to be
\begin{eqnarray}\label{grnfnc}
G^{\rm R}_{t~t}(m_q, \mu_d, \sigma) &=& - \frac{1}{g_5^2}
\left( \int^0_1 du \frac{1}{e^{-\Phi}\sqrt{g}g^{uu}g^{tt}} + D_1 \right)^{-1} + {\cal O}(\omega^2,q^2),
\end{eqnarray}
where $A^{(0)}_t$ is normalized to $1$.
By taking $q \to 0$ with $\omega=0$, we can obtain the quark number susceptibility $\chi_q$ through eq.(\ref{suss}).
In the following computation, we will neglect
the high order correction part of ${\cal O}(\omega^2,q^2)$, and also take $g_5$ to be $g_5 =1$ for simplicity as the physics property of the green function $G^{\rm R}_{t~t}$ is irrelevant to its absolute value. 

We shall compute  $\chi_q/T^2$ as function of three parameters
$\sigma$, $m_q$ and $\mu_d$. Typically, we will discuss the
variation of $\sigma$ with a range $\pm 25$ toward the value of
$\sigma^{1/3}$ in Table.\ref{parameter}. This does not mean the
uncertainty but the range to see how the QNS $\chi_q/T^2$ changes as
the dynamical quark mass. On the other hand, we shall consider the
uncertainty of about $10 \% $ toward the values of $\mu_d$ in
Table.\ref{parameter}, which is mainly caused from the difference
between the predicted mass spectra and experimental data for the
resonance meson states\cite{Sui:2009xe}.
%
%
\begin{table}[h]
\begin{center}
\begin{tabular}{rc}
 Ia : & $\sigma^{1/3} = 275 \pm 25$ ($m_q \simeq 4.16 \mp 1.2$), $\gamma = 0.178$ and $\mu_d = 445$ ($\mu_g =363$)\\
 Ib : & $\sigma^{1/3} = 265 \pm 25$ ($m_q \simeq 4.64 \mp 1.2$), $\gamma = 0.178$ and $\mu_d = 445$ ($\mu_g =257$)\\
IIa : & $\sigma^{1/3} = 265 \pm 25$ ($m_q \simeq 4.44 \mp 1.2$), $\gamma = 0.178$ and $\mu_d = 445$ ($\mu_g =363$)\\
IIb : & $\sigma^{1/3} = 272 \pm 25$ ($m_q \simeq 4.07 \mp 1.2$), $\gamma = 0.178$ and $\mu_d = 445$ ($\mu_g =257$)
\end{tabular}
\caption{The parameters used in Fig.\ref{Figa}. $\gamma$ is dimensionless and the unit of the rest are MeV. $m_q$ is automatically fixed by the relation.(\ref{GMOR}).
}
\label{prmtFiga}
\end{center}
\end{table}

\begin{table}[h]
\begin{center}
\begin{tabular}{rc}
 Ia : & $m_q = 4.16$, $\sigma^{1/3} = 275$, $\gamma = 0.178$ and $\mu_d = 445 \pm 50$ ($\mu_g \simeq 363 \mp 40$)\\
 Ib : & $m_q = 4.64$, $\sigma^{1/3} = 265$, $\gamma = 0.136$ and $\mu_d = 445 \pm 50$ ($\mu_g \simeq 257 \mp 30$)\\
IIa : & $m_q = 4.44$, $\sigma^{1/3} = 265$, $\gamma = 0.153$ and $\mu_d = 445 \pm 50$ ($\mu_g \simeq 363 \mp 40$)\\
IIb : & $m_q = 4.07$, $\sigma^{1/3} = 272$, $\gamma = 0.112$ and $\mu_d = 445 \pm 50$ ($\mu_g \simeq 257 \mp 30$)
\end{tabular}
\caption{The parameters used in Fig.\ref{Figb}.
$\gamma$ is dimensionless and the unit of the rest are MeV.
The relation for $\mu_d^2$ and $\mu_g^2$ are given in eq.(\ref{vinft}).}
\label{prmtFigb}
\end{center}
\end{table}

\begin{figure}[h!]
\begin{center}
\includegraphics[width=46mm,clip]{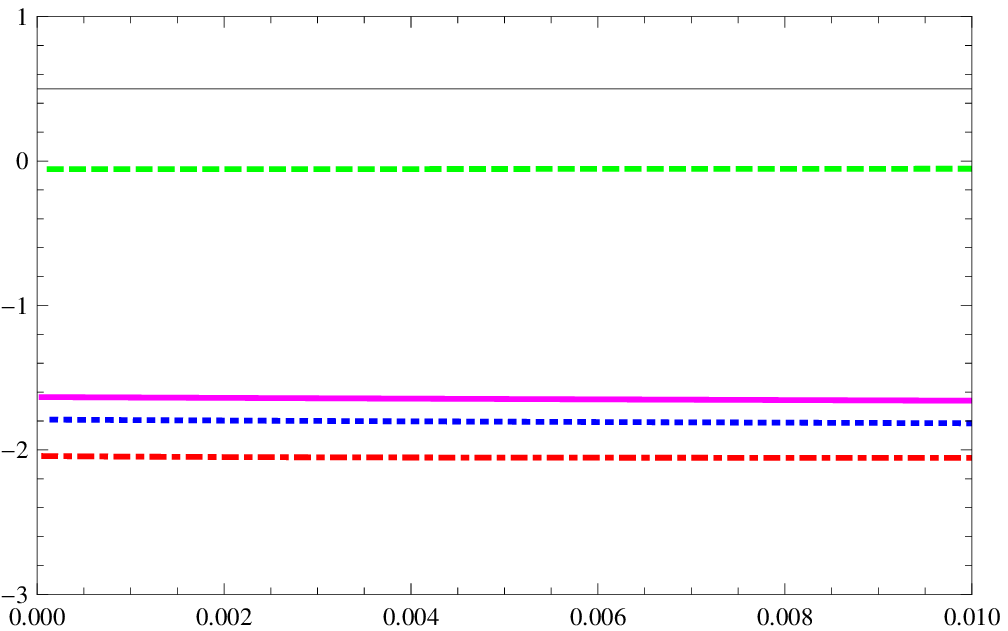}
\includegraphics[width=46mm,clip]{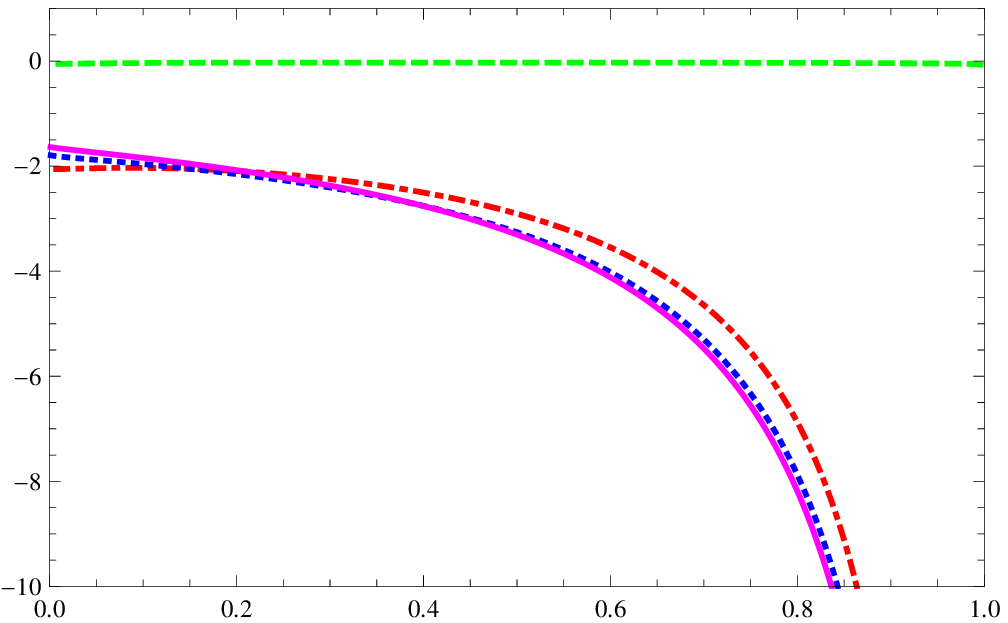}
\includegraphics[width=46mm,clip]{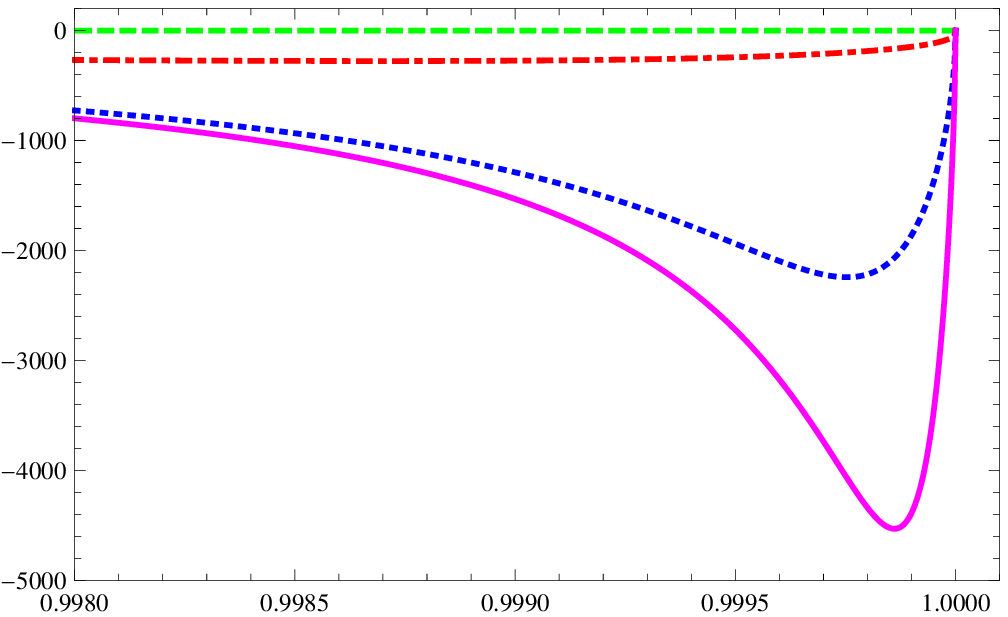}
\end{center}
\caption{ The figures show the regularity of the solution $A_t'$ in
the entire bulk. Note the scale taken in each figures. The x-axis
and y-axis denote the radial coordinate $u$ and the real part of the
solution $A_t'$ in the Ia model, respectively.
On the other hand, the imaginary-part in all models always takes the value extremely close to zero.
Considering it as just a numerical error, we then skip to show it here.
The three figures are for the real-part of the
model Ia: the left one indicates no-UV divergence, the center one
shows the behavior in the whole region, and the right one shows
finiteness and the boundary condition is taken correctly around the
horizon. The behaviors for other models are the same. In the
figures, we have taken $u \in [10^{-25},1-10^{-10}]$ with the
in-going boundary condition, and $q$ and $\omega$ are taken to be zero.
The parameters used here are given in Table.\ref{prmtFigb} with
temperature changing as T=120 (green, dashed), T=200 (red, dot-dashed), T=300 (blue, dotted),
T=400 (magenta, solid), (Unit is MeV).
}\label{FigdAcnt}
\end{figure}

Before making the computation of quark number susceptibility,
we may check that the solutions $A_t'$ are well-defined in the entire bulk (Fig.\ref{FigdAcnt}).

\newpage

\begin{figure}[h]
\begin{center}
\includegraphics[width=55mm,clip]{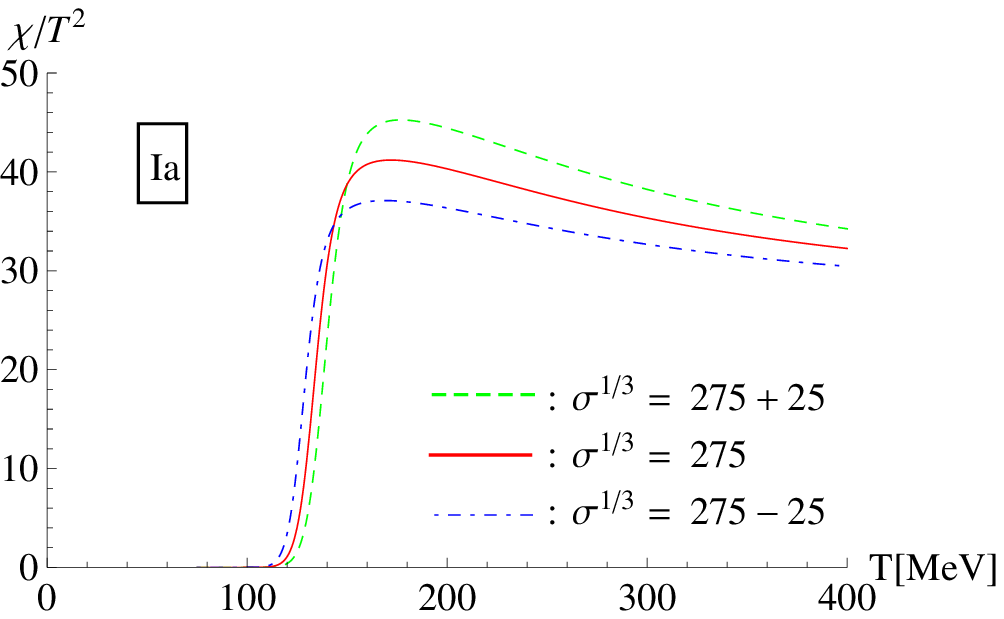}
\includegraphics[width=55mm,clip]{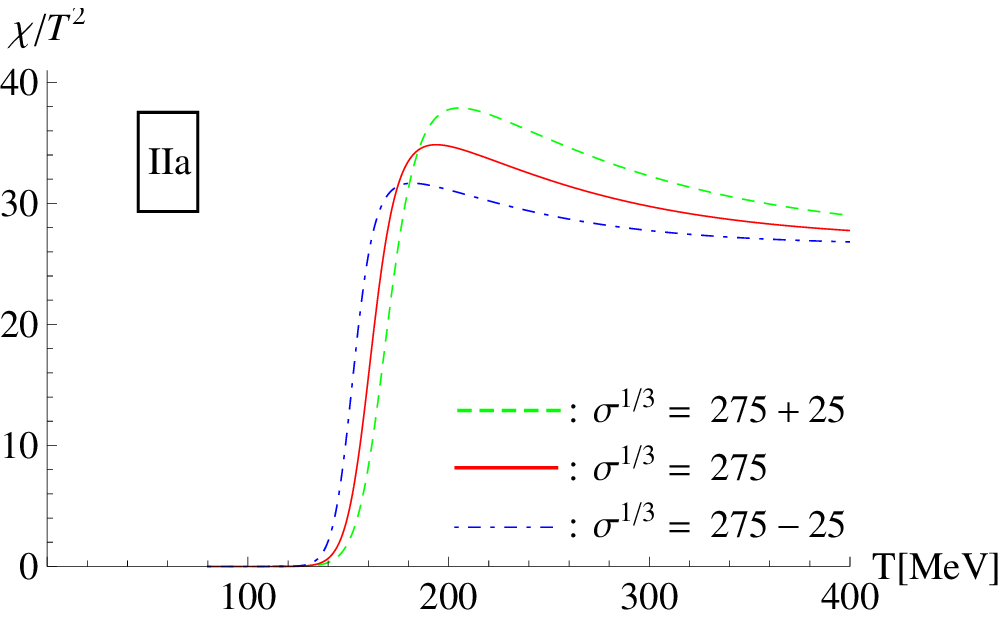}
\\
\includegraphics[width=55mm,clip]{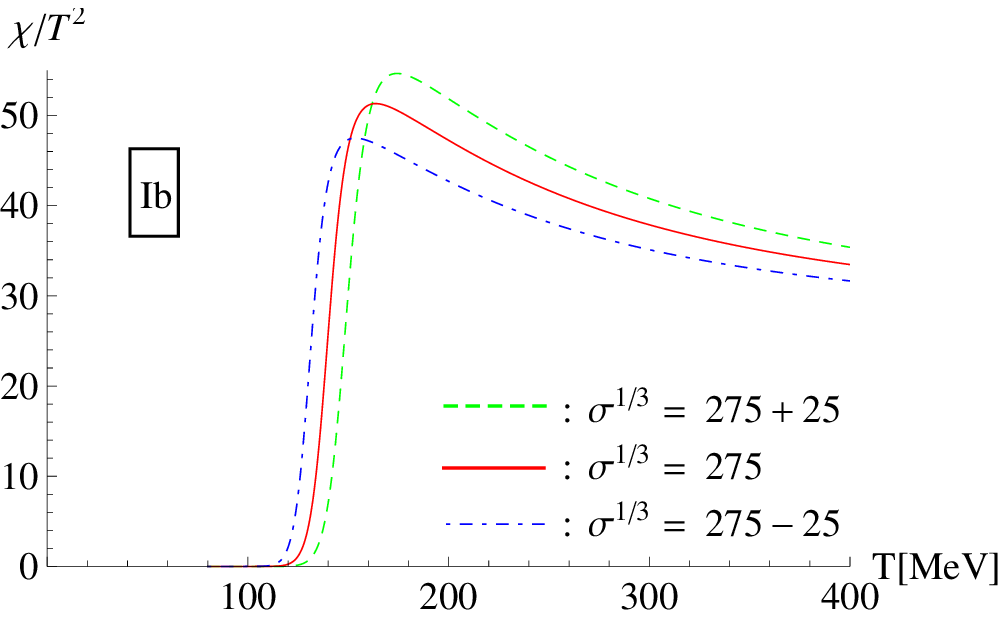}
\includegraphics[width=55mm,clip]{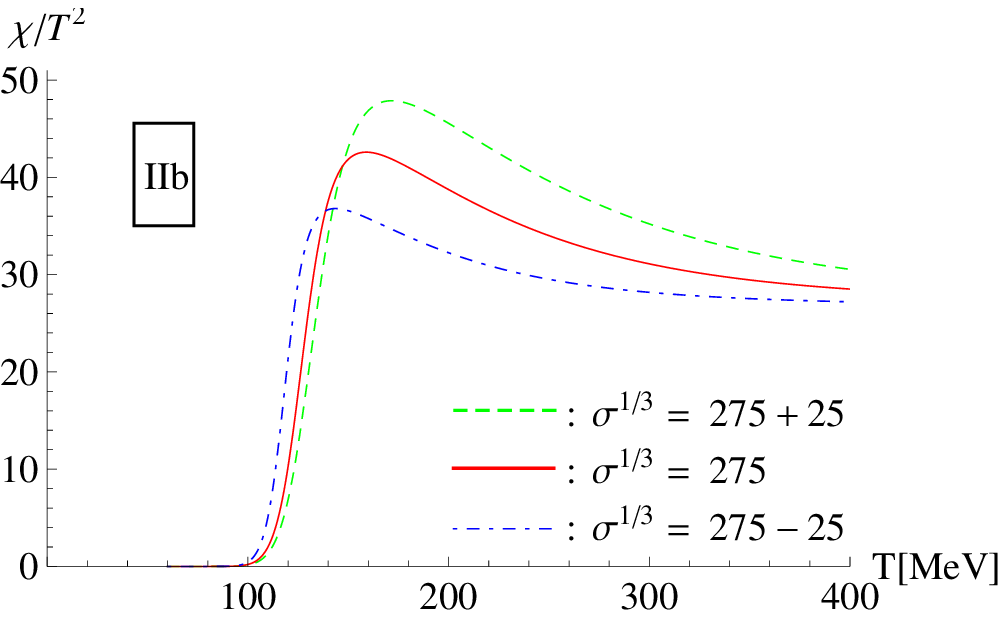}
\end{center}
\caption{
Results with fixing $\mu_d$ and varying $m_q$ and $\sigma^{1/3}$ as in Table.\ref{prmtFiga}.
The different curves from top to bottom in all figures correspond to $\sigma^{1/3}=275+25$ (green, dashed), $\sigma^{1/3}=275$ (red, solid) and $\sigma^{1/3}=275-25$ (blue, dot-dashed). The changes of $\sigma$ is just to see the dependence of QNS on the constituent quark mass given by $\bar{m}_q=m_q+\sigma/\mu^2$. It is observed that the appearing of peaks and its locations in all models agree with the results of lattice QCD \cite{AlltonEjiri}.
}\label{Figa}
\end{figure}

\begin{figure}[h]
\begin{center}
\includegraphics[width=55mm,clip]{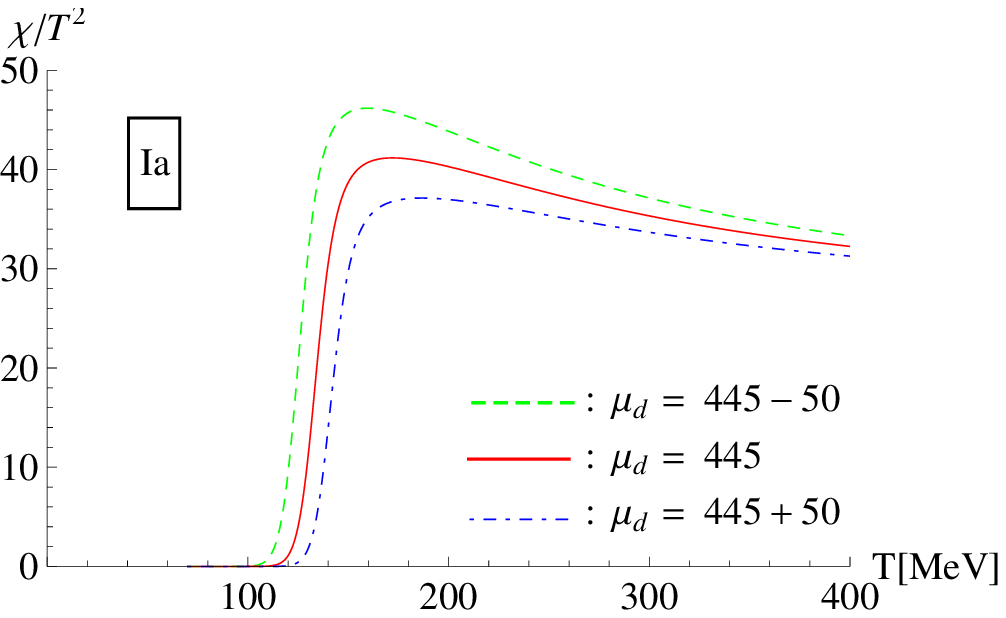}
\includegraphics[width=55mm,clip]{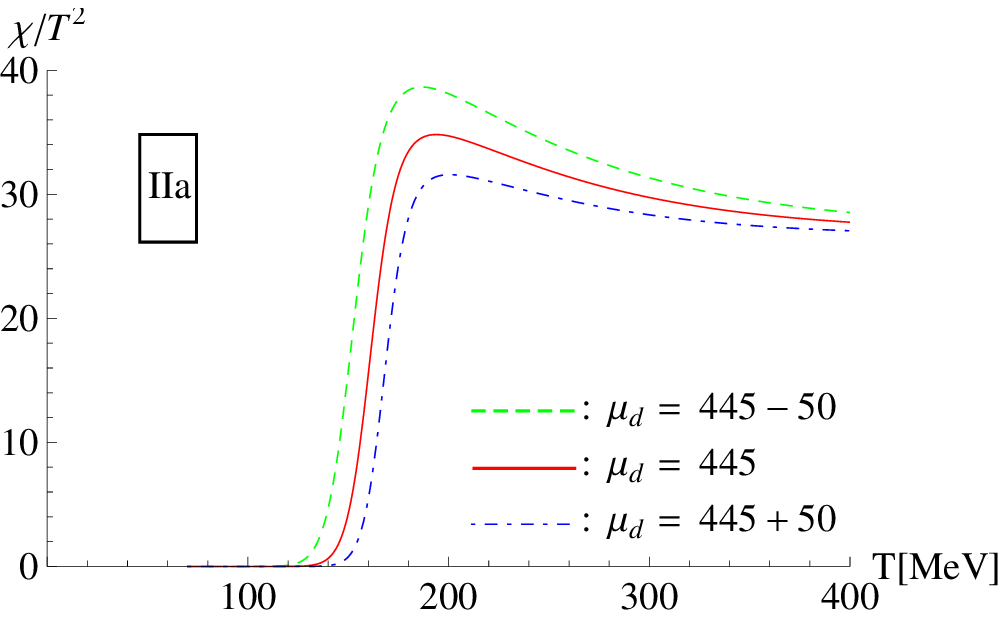}
\\
\includegraphics[width=55mm,clip]{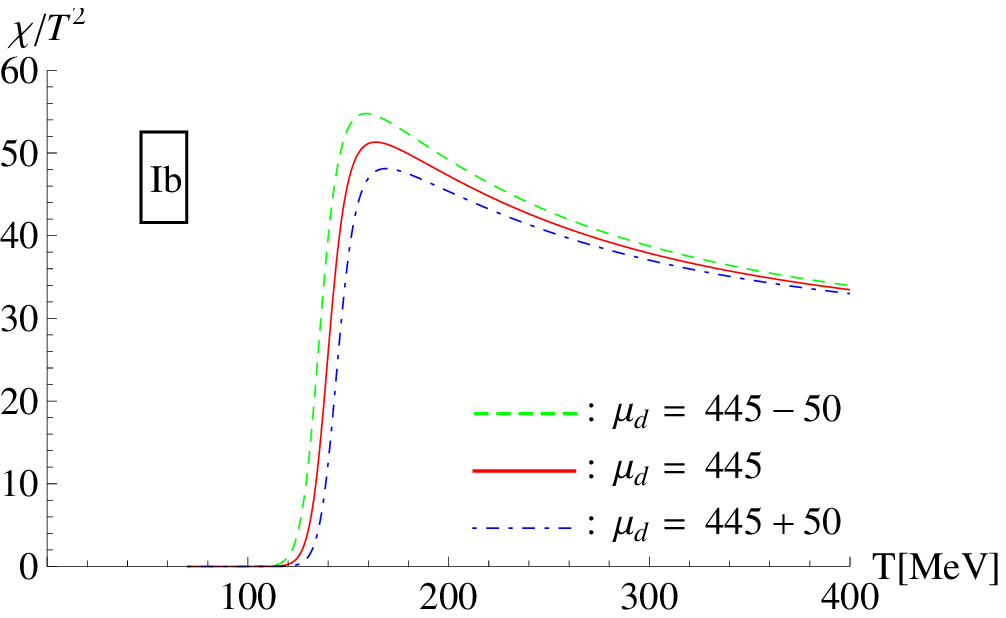}
\includegraphics[width=55mm,clip]{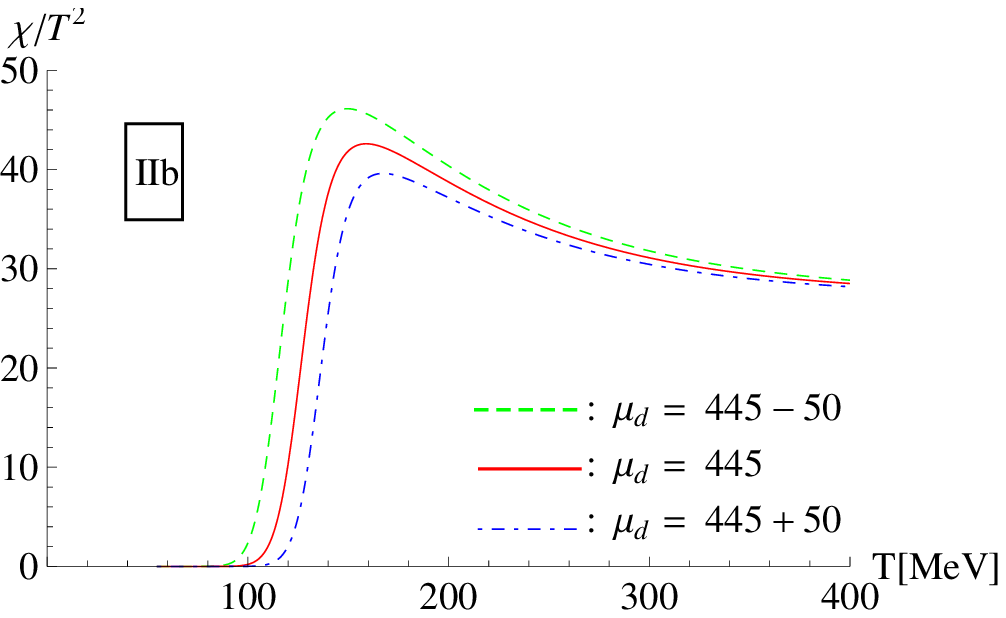}
\end{center}
\caption{
Results with fixing $m_q$ and $\sigma^{1/3}$ as in Table.\ref{prmtFigb} and varying $\mu_d$.
The different curves from top to bottom in all figures correspond to $\mu_d=445-50$(green, dashed), $\mu_d=445$(red, solid) and $\mu_d=445+50$(blue, dot-dashed).
The possible differences with regard to $\mu_d$ arise from the one between the predicted resonance meson mass spectra and the experimental data. It can be seen that the appearing of peaks and its locations in all models are in good agreement with the lattice QCD' results. \cite{AlltonEjiri}.
}\label{Figb}
\end{figure}

\noindent
From the results shown in Fig.\ref{FigdAcnt}, it is seen that there is no divergence involved in our present considerations,
therefore we do not need to care the regularization for the retarded green function \cite{ebs-qns1,ebs-hydro1}.
Thus by using the parameters in Table.\ref{prmtFiga} and \ref{prmtFigb}, we obtain the results shown in Fig.\ref{Figa} and \ref{Figb}.


%



It can be seen from Figs.\ref{Figa} and \ref{Figb} that the QNS $\chi_q/T^2$ emerges rapidly with a continuous blow-up in a narrow
temperature interval as the temperature increases. There is a peak when the temperature reaches to be around $T \sim 160 \sim 190$ MeV depending slightly on models, and after that, the behaviors of $\chi_q/T^2$ get slowly in all the models. As mentioned in the introduction, the results of lattice QCD \cite{Mc87, AlltonEjiri} have shown that when the phase changes from the hadron phase to the quark-gluon-plasma phase at zero chemical potential, $\chi_q/T^2$ grows rapidly and continuously with a peak around $T \simeq 170$ MeV. Therefore, the appearing of peaks and its locations in all models are in good agreement with the lattice QCD' results

Let us now make a determination for the critical temperature in each model, which is simply fixed at the location of the peak
\begin{eqnarray}\label{Tc}
 \textrm{Ia :}~ \quad T_c  &=& 173.6 \pm  3.2\big|_{\sigma^{1/3}} \pm 11.7\big|_{\mu_d},\\
 \textrm{Ib :}~ \quad T_c  &=& 164.4 \pm  8.1\big|_{\sigma^{1/3}} \pm  5.2\big|_{\mu_d},\\
\textrm{IIa :}~ \quad T_c  &=& 194.9 \pm  7.4\big|_{\sigma^{1/3}} \pm  9.2\big|_{\mu_d},\\
\textrm{IIb :}~ \quad T_c  &=& 159.5 \pm  7.8\big|_{\sigma^{1/3}} \pm  9.4\big|_{\mu_d} ,
\end{eqnarray}
where the uncertainties come from the quantities $|_{\sigma^{1/3}}$ and $|_{\mu_d}$ as denoted by the subindices, the possible variations of $|_{\sigma^{1/3}}$ and $|_{\mu_d}$ are given in Tables\ref{prmtFiga} and \ref{prmtFigb}.

\begin{figure}[h]
\begin{center}
\includegraphics[width=54mm,clip]{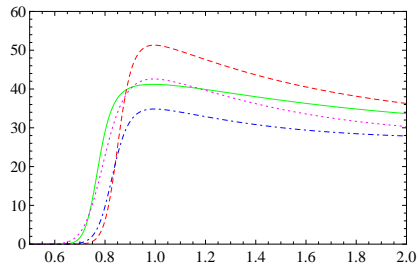}
\end{center}
\caption{
The results of all models with the parameters given in Table.\ref{parameter}.
The curves corresponding to different models may be distinguished from their locations corresponding to large value of $T/T_c\geq 1.5$,
form top to bottom, they are: Ib (red, dashed), Ia (green, solid), IIb (Magenta, dashed), IIa (blue, dot-dashed). The x-axis and y-axis denote the normalized
temperature $T/T_c$ and $\chi/T^2$, respectively.
}\label{FigTc}
\end{figure}

From the above analysis and results, we would like to address the following points:
First, we may draw attention to the effect of the VEV $v(z)$. From Fig.\ref{Figa},
it is seen that the effects in all models are similar as long as it satisfies the boundary conditions (\ref{vinft}) and (\ref{vzero}) with the given $\alpha$,
regardless of the form of $v(z)$.
On the other hand, it turns out that the appearing of the peaks moves to high temperature as $\sigma^{1/3}$ increases.
Next, let us turn to the effects of $\mu_g^2$.
From Fig.\ref{Figb},
it is noticed that the behaviors of the peak become softer in all models, and the locations of the peak shift to the high temperature as $\mu_g^2$ increases.
As for the effect of quark mass, considering the constituent quark mass defined via $\bar{m}_q = m_q + \sigma/\mu^2$ with $m_q$ and $\sigma/\mu^2$ denoting the current quark mass and the dynamically generated quark mass via chiral condensation, respectively.
As the relation eq.(\ref{GMOR}) is kept to be held, the current quark mass $m_q$ and the dynamically generated quark mass or chiral condensation $\sigma$ are varying oppositely, which can be seen explicitly from Table.\ref{prmtFiga}. It can also be seen from Fig.\ref{Figa} that when the constituent quark mass $\bar{m}_q$ is going to be increased, which is toward the increased $\sigma$ associated with the decreased $m_q$ due to the relation eq.(\ref{GMOR}), the behavior of the QNS is getting steeper, and vice versa. This is consistent with the general point of view that the more massive the current quark mass becomes, the more moderate the behavior of the peak for the QNS $\chi/T^2$ gets.
In this sense, it can be interpreted that the effect of current quark mass $m_q$ is more effective
than the one of dynamically generated quark mass or the chiral condensation $\sigma$.
%
Finally, we plot in Fig.\ref{FigTc} the results in all models with parameters given in Table.\ref{parameter}
and the temperature is normalized by the critical temperature $T_c$, so that each peak denoting $T_c$ come to $1$ on the x-axis.


\section{Summary and Remarks}\label{chap:Summary}

In this paper, we have extended the predictive soft-wall AdS/QCD model\cite{Sui:2009xe} to a finite temperature system and carried out the calculation for the quark number susceptibility with finite quark mass. It has been shown that the critical behaviors in all models agree well with the lattice QCD calculation\cite{Mc87, AlltonEjiri}. This further demonstrates that the predictive soft-wall AdS/QCD model proposed in\cite{Sui:2009xe} is a reasonable and realistic effective model for describing the QCD confinement and chiral symmetry breaking, where the conformal symmetry breaking at infrared region due to nonperturbative QCD effects is characterized by the modified infrared behavior of metric parameterized by $\mu_g^2$ in the metric and the nontrivial dilaton background (namely the soft-wall) which is determined by both the IR parameter $\mu_g^2$ and the VEV $v(z)$. It seems reasonable to suppose that $v(z)$ has almost the same effects as long as it satisfies the boundary condition (\ref{vinft}) and (\ref{vzero}). In addition,
it has been noticed that the critical temperature is put to a low temperature by the soft-wall,
while the effect of IR behavior $\mu_g^2$ will pull it up to high temperature.
Thus the matching of the critical behaviors can be seen as the result of a very fine balance between $\mu_g^2$ and the soft-wall.
As for the effect of quark mass given by the constituent quark mass $\bar{m}_q=m_q + \sigma/\mu^2$, it has been shown that the effect of current quark mass $m_q$ is more significant than the one of chiral condensation $\sigma$ or dynamically generated quark mass when the relation eq. (\ref{GMOR}) is imposed, which may be explicitly seen from the relation $\bar{m}_q = m_q +
f_{\pi}^2 m_{\pi}^2/(2m_q \mu^2)$.

Here we only consider two flavors, it would be interesting to investigate the nature of QCD transition with three flavor case, and also with including the chemical potential.  On the other hand, the quartic interaction term of the scalar field has been dropped in eq.(\ref{lagran}),
while it has been shown that it has a sizable role in the computation of the mass spectra of resonance mesons\cite{Sui:2009xe}. Also the back reaction effects of the dilaton field and gravity may be the core in the improvement of our model\cite{SWWX}. We shall consider those interesting topics in our future studies.

\vspace*{5mm}

\noindent{\large{\bf Acknowledgments}}

\vspace*{2mm}
\noindent
The authors would like to thank Y.B. Yang for useful discussions.
This work is supported in part the National
Nature Science Foundation of China (NSFC) under Grants No. 10975170,
No. 10821504 and No. 10905084; and the Project of Knowledge Innovation
Program (PKIP) of the Chinese Academy of Science.
\appendix

\end{document}